\documentclass[aps,prd,eqsecnum,preprint,tightenlines,nofootinbib,showpacs]{revtex4-1}
\usepackage{graphicx,amsmath,latexsym}

\newcommand{\bea}{\begin{eqnarray}}
\newcommand{\eea}{\end{eqnarray}}
\newcommand{\be}{\[}
\newcommand{\ee}{\]}
\newcommand{\ub}[1]{\underline{#1}}
\newcommand{\ob}[1]{\overline{#1}}
\newcommand{\Pminus}{{\cal P}^-}

\begin{document}

\title{A light-front coupled-cluster method for quantum field theories\footnote{Presented 
at the Sixth International Conference on Quarks and Nuclear Physics,
		 April 16-20, 2012,
		 Ecole Polytechnique, Palaiseau, Paris.  To appear in the proceedings.}}

\author{John R. Hiller}
\affiliation{Department of Physics \\
University of Minnesota-Duluth \\
Duluth, Minnesota 55812 USA}

\date{\today}

\begin{abstract}
The Hamiltonian eigenvalue problem for bound states of a quantum
field theory is formulated in terms of Dirac's light-front coordinates
and then approximated by the exponential-operator technique of
the standard coupled-cluster method.  This approximation eliminates
any need for the usual approximation of Fock-space truncation.  
Instead, the exponential operator is truncated and the terms retained
are determined by a set of nonlinear integral equations.  These
equations are solved simultaneously with an effective eigenvalue
problem in the valence sector, where the number of constituents
is small.  Matrix elements can be calculated, with extensions of
techniques from standard coupled-cluster theory.
\end{abstract}

\maketitle

\section{Introduction}

The light-front coupled-cluster (LFCC) method~\cite{LFCClett}
is intended as a method for the calculation of
hadron structure in terms of Fock-state wave functions
without the usual truncation of Fock space.  The Fock-state
representation of a hadron, such as a proton, takes the form
\be
|p\rangle=\psi_{uud}|uud\rangle +\psi_{uudg}|uudg\rangle 
+\psi_{uudgg} |uudgg\rangle+\psi_{uudq\bar{q}}|uudq\bar{q}\rangle+\cdots.
\ee
This is to be an eigenstate of the QCD Hamiltonian
\be
\left(\mbox{K.E.}+V_{\rm QCD}\right)|p\rangle=E_p|p\rangle
\ee
with $E_p=\sqrt{m_p^2+p^2}$ and $V_{\rm QCD}$ the 
interaction terms of gluon emission and absorption
by quarks, quark-antiquark production and annihilation,
and the three and four-gluon vertices.
This is equivalent to a set of coupled integral equations
for the wave functions.

Usually, this infinite set of equations is made finite
by truncation of Fock space; however, this introduces
uncanceled divergences~\cite{SecDep}.
For example, the Ward identity of gauge theories is
destroyed by truncation, because the limitation on
the number of particles allowed in intermediate states
removes some of the contributions to the identity.
There is a direct analog in Feynman perturbation theory,
where separation of a covariant diagram into
time-ordered diagrams, and removal of those
that include intermediate states
with more particles than some finite limit,
destroys covariance, disrupts regularization, and 
induces spectator dependence for subdiagrams.
In the nonperturbative case, this happens not just
to some finite order in the coupling but to all orders.

The method is formulated in terms of light-cone
coordinates~\cite{Dirac,DLCQreview}.
The time coordinate is $x^+=t+z$, and the spatial
coordinates are $\ub{x}=(x^-,\vec{x}_\perp)$, with
$x^-\equiv t-z$ and $\vec{x}_\perp=(x,y)$.  The
light-cone energy is $p^-=E-p_z$, and the momentum is
$\ub{p}=(p^+,\vec{p}_\perp)$, with 
$p^+\equiv E+p_z$ and $\vec{p}_\perp=(p_x,p_y)$.
This leaves the mass-shell condition
$p^2=m^2$ as $p^-=\frac{m^2+p_\perp^2}{p^+}$
and the mass eigenvalue problem as~\cite{PauliBrodsky}
\be
\Pminus|\ub{P}\rangle=\frac{M^2+P_\perp^2}{P^+}|\ub{P}\rangle, \;\;\;\;
\underline{\cal P}|\ub{P}\rangle=\underline{P}|\ub{P}\rangle.
\ee

The advantages of this coordinate choice include
the absence of spurious vacuum contributions to eigenstates
and a boost-invariant separation of internal and external momenta.
Vacuum contributions are suppressed because $p^+$ is
positive for all particles; the vacuum cannot produce particles
and conserve $p^+$.  The separation of internal momenta is
obtained by defining longitudinal momentum fractions
$x_i\equiv p_i^+/P^+$ and relative transverse momenta 
$\vec k_{i\perp}\equiv \vec p_{i\perp}-x_i\vec P_\perp$.

The original coupled-cluster (CC) method was developed for
the many-body Schr\"odinger equation in nuclear physics~\cite{CCorigin}
and later applied to the many-electron problem in molecules~\cite{Cizek}.
The method has become an important tool in nuclear physics
and physical chemistry for the $N$-body problem in nonrelativistic
quantum systems~\cite{CCreviews}.  The basic idea is
to form an eigenstate as $e^T|\phi\rangle$, where
$|\phi\rangle$ is a product of single-particle states
and the terms in $T$ annihilate states in $|\phi\rangle$
and create excited states, to build in correlations;
however, the number of particles does not change.
The approximation made is to truncate $T$ at some number
of excitations.  The LFCC
method uses the mathematics of these constructions but
applies it to a situation where $|\phi\rangle$ contains
a small number of particles, $T$ adds additional
particles, and the states are eigenstates of momentum
with Dirac-delta normalization.

\section{Light-front coupled-cluster method}

To solve the fundamental eigenvalue problem
$\Pminus|\psi\rangle=\frac{M^2+P_\perp^2}{P^+}|\psi\rangle$
by the LFCC method~\cite{LFCClett}, we write the eigenstate as 
$|\psi(\ub{P})\rangle=\sqrt{Z}e^T|\phi(\ub{P})\rangle$ and seek
solutions for the valence state $|\phi(\ub{P})\rangle$
and the operator $T$.  This operator contains
terms that only increase particle number, while
conserving $J_z$, light-front momentum $\ub{P}$,
charge, and the other quantum numbers of the eigenstate.
The constant $Z$ controls normalization, which is chosen to be
$\langle\psi(\ub{P}')|\psi(\ub{P})\rangle=\delta(\ub{P}'-\ub{P})$,
with the valence state normalized in the same way:
$\langle\phi(\ub{P}')|\phi(\ub{P})\rangle=\delta(\ub{P}'-\ub{P})$.
Because $p^+$ is positive, $T$ must include annihilation,
and powers of $T$ include contractions.
This converts the original eigenvalue problem into
an eigenvalue problem for the valence state
\be
P_v\ob{\Pminus}|\phi(\ub{P})\rangle=\frac{M^2+P_\perp^2}{P^+}|\phi(\ub{P})\rangle,
\ee
with $\ob{{\cal P}-}=e^{-T} \Pminus e^T$ the effective Hamiltonian
and $P_v$ the projection onto the valence Fock sector,
and into an auxiliary equation for $T$
\be
(1-P_v)\ob{\Pminus}|\phi(\ub{P})\rangle=0.
\ee

Calculation of expectation values, and more generally matrix elements,
requires some care, to avoid any necessity of computing the infinite
sum implied by the inner product $\langle\phi|e^{T^\dagger}e^T|\phi\rangle$.
This can be done, with use of constructions from the CC method~\cite{CCreviews}.
The expectation value 
\be
\langle\hat O\rangle=\frac{\langle\phi| e^{T^\dagger}\hat O e^T|\phi\rangle}
                      {\langle\phi| e^{T^\dagger} e^T|\phi\rangle}
\ee
for an operator $\hat{O}$ is rewritten as 
$\langle\hat O\rangle=\langle\tilde\psi|\ob{O}|\phi\rangle$
in terms of an effective 
operator $\ob{O}\equiv e^{-T}\hat O e^T$ and a left eigenvector of $\ob{\Pminus}$:
\be
\langle\tilde\psi|=\langle\phi|\frac{e^{T^\dagger}e^T}
      {\langle\phi|e^{T^\dagger} e^T|\phi\rangle}.
\ee
The effective
operator can be computed from the Baker--Hausdorff expansion
$\ob{O}=\hat O + [\hat O,T]+\frac12[[\hat O,T],T]+\cdots$.
The bra $\langle\tilde\psi|$ is seen to be a left eigenstate by the following
steps:
\be
\langle\tilde\psi|\ob{\Pminus}
=\langle\phi|\frac{e^{T^\dagger}\Pminus e^T}{\langle\phi| e^{T^\dagger} e^T|\phi\rangle}
=\langle\phi|\ob{\Pminus}^\dagger \frac{e^{T^\dagger}e^T}
                            {\langle\phi| e^{T^\dagger} e^T|\phi\rangle}
=\frac{M^2+P_\perp^2}{P^+}\langle\tilde\psi|.
\ee
Also, it has the normalization
\be
\langle\tilde\psi(\ub{P}')|\phi(\ub{P})\rangle
=\langle\phi(\ub{P}')|\frac{e^{T^\dagger}e^T}{\langle\phi| e^{T^\dagger} e^T|\phi\rangle}|\phi(\ub{P})\rangle
=\delta(\ub{P}'-\ub{P}).
\ee

As formulated to this point, the new valence eigenvalue problem and the
auxiliary condition for $T$ provide an exact representation of the 
original eigenvalue problem.  The LFCC method then invokes a truncation,
not of Fock space but of the terms in $T$ and in the projection
$1-P_v$.  This leads to a finite
set of auxiliary equations for a finite set of functions that define
the truncated $T$ operator and to a finite number of terms in the
valence projection of the effective Hamiltonian $\ob{\Pminus}$.
The latter is conveniently expressed in terms of a Baker--Hausdorff
expansion where the number of contributing terms is finite.  Similarly,
the contributions to matrix elements of operators are also finite in number.

\section{Application to a soluble model}

To illustrate the method, we apply it~\cite{LFCClett}
to an exactly soluble model~\cite{GreenbergSchweber},
a light-front analog of the Greenberg--Schweber model
with a static fermionic source that emits and absorbs bosons
without changing its spin.  The model lacks full covariance
because of the static source; all states are limited to having
a fixed total transverse momentum $\vec P_\perp$, 
which we take to be zero.  In this context, not
all features of the method are apparent, but the model
is sufficient to show how the method is used.

The light-front Hamiltonian is~\cite{LFCClett}
\bea \label{eq:Pminus}
\lefteqn{\Pminus = \int d\ub{p} \frac{M^2+M'_0 p^+}{P^+}\sum_s b_s^\dagger(\ub{p})b_s(\ub{p})
  +\int d\ub{q}\sum_l(-1)^l \frac{\mu_l^2+q_\perp^2}{q^+} a_l^\dagger(\ub{q})a_l(\ub{q})}&& \\
 &+\frac{g}{P^+} \int \frac{d\ub{p}d\ub{q}}{\sqrt{16\pi^3 q^+}}
  \sum_{ls}\left(\frac{p^+}{p^++q^+}\right)^\gamma 
  \left[a_l^\dagger(\ub{q})b_s^\dagger(\ub{q})b_s(\ub{p}+\ub{q})
    +b_s^\dagger(\ub{p}+\ub{q})b_s(\ub{p})a_l(\ub{q})\right], \nonumber
\eea
where $a_0^\dagger$ creates a ``physical'' boson of mass $\mu_0$,
$a_1^\dagger$ creates a Pauli--Villars (PV) boson of mass $\mu_1$,
and $b_s^\dagger$ creates the fermion with mass $M$ and spin $s$.
The valence state is the bare-fermion state
$|\phi^\sigma(\ub{P})\rangle=b_\sigma^\dagger(\ub{P})|0\rangle$.
The $T$ operator is truncated to a single boson emission:
\be
T=\sum_{ls}\int d\ub{q} d\ub{p} \, t_{ls}(\ub{q},\ub{p})
   a_l^\dagger(\ub{q})b_s^\dagger(\ub{p})b_s(\ub{p}+\ub{q}),
\ee
and the projection $1-P_v$ is
truncated to the one-fermion/one-boson sector.

This form generates the exact solution, with
\be
t_{ls}(\ub{q},\ub{p})=\frac{-g}{\sqrt{16\pi^3 q^+}}\left(\frac{p^+}{p^++q^+}\right)^\gamma
   \frac{q^+/P^+}{\mu_l^2+q_\perp^2}.
\ee
The fermion self-energy contribution $M'_0$ is the same
in all Fock sectors and the effective Hamiltonian~\cite{LFCClett}
contains all three of the contributions analogous to
those for the Ward identity in QED.
The fact that the self-energy loop is the same in the valence
sector and the one-fermion/one-boson sector plays a critical role
in yielding the exact solution.

To compute an observable, we consider the Dirac form factor
for the dressed fermion.  It can be obtained from
a matrix element of the current 
$J^+=\ob{\psi}\gamma^+\psi$ coupled to a photon of momentum $q$.
The matrix element is generally~\cite{BrodskyDrell}
\be
\langle\psi^\sigma(\ub{P}+\ub{q})|16\pi^3J^+(0)|\psi^\pm(\ub{P})\rangle
=2\delta_{\sigma\pm}F_1(q^2)\pm\frac{q^1\pm iq^2}{M}\delta_{\sigma\mp}F_2(q^2),
\ee
with $F_1$ and $F_2$ the Dirac and Pauli form factors.
In the model, the fermion cannot flip its spin; therefore,
$F_2$ is zero, and we compute only $F_1$.  Also, there
are no contributions from fermion-antifermion pairs, so that
the current is simply
\be
J^+(0)=2\sum_s\int\frac{d\ub{p}'}{\sqrt{16\pi^3}}\int\frac{d\ub{p}}{\sqrt{16\pi^3}}
    b_s^\dagger(\ub{p}')b_s(\ub{p}).
\ee

In the LFCC method, the form factor is approximated by the matrix element
\be
F_1(q^2)=8\pi^3\langle\widetilde\psi^\pm(\ub{P}+\ub{q})|\ob{J^+(0)}|\phi^\pm(\ub{P})\rangle,
\ee
with $\ob{J^+(0)}=J^+(0)+[J^+(0),T]+\cdots$.
The truncated left-hand eigenvector is
\be
\langle\widetilde\psi^\sigma(\ub{P})|
       =\langle\phi^\sigma(\ub{P})|
+\sum_{ls}\int d\ub{q}\theta(P^+-q^+)
l_{ls}^{\sigma*}(\ub{q},\ub{P})\langle0|a_l(\ub{q}) 
b_s(\ub{P}-\ub{q}),
\ee
where $l_{ls}^\sigma$ is the 
left-hand one-fermion/one-boson wave function.
If this wave function is assumed to take the form
\be
l_{ls}^\sigma(\ub{q},\ub{P})=\delta_{\sigma s}\frac{-g}{\sqrt{16\pi^3 q^+}}
   \left(\frac{P^+-q^+}{P^+}\right)^\gamma \frac{q^+/P^+}{\mu_l^2+q_\perp^2}\tilde{l}(q^+/P^+),
\ee
substitution into the left-hand eigenvalue problem yields a one-dimensional
integral equation for $\tilde{l}(y)$
\be
\tilde{l}(y)=1+\frac{g^2}{16\pi^2}\frac{\mu_1^2-\mu_0^2}{\mu_0^2\mu_1^2}
\int_0^1 dy' (1-y')^{2\gamma} y'[(1-y)^2\tilde{l}(y'(1-y))-\tilde{l}(y')].
\ee
The solution of this equation can then be used to compute matrix elements.

Only the first two terms of the Baker--Hausdorff expansion of
$\ob{J^+(0)}$ contribute to the matrix element.  The first term 
contributes $1/8\pi^3$ and second contributes
\bea
\lefteqn{\langle\widetilde\psi^\pm(\ub{P}+\ub{q})|[J^+(0),T]|\phi^\pm(\ub{P})\rangle
=\frac{1}{8\pi^3}\sum_{l}(-1)^l \int d\ub{q}'\theta(P^++q^+-q^{\prime +})} && \nonumber \\
&& \times l_{l\pm}^\pm(\ub{q}',\ub{P}+\ub{q})
  [\theta(P^+-q^{\prime +})t_{l\pm}(\ub{q}',\ub{P}-\ub{q}')
    -t_{l\pm}(\ub{q}',\ub{P}+\ub{q}-\ub{q}')].
\eea
The form factor is then
\bea \label{eq:formfactor}
F_1(q^2)&=&1+\frac{g^2}{16\pi^2}(1+\alpha)\frac{\mu_1^2-\mu_0^2}{\mu_0^2\mu_1^2}
\left[\int_0^{1/(1+\alpha)} dy\,\tilde{l}(y) y(1-y)^\gamma[1-(1+\alpha)y]^\gamma
                     \right. \nonumber \\
&& \rule{1.5in}{0mm} \left.
-\int_0^1 dy\,\tilde{l}(y) y(1-y)^{2\gamma}\right], 
\eea
with $\alpha\equiv q^+/P^+$.
In the limit of $q^2\rightarrow0$,
we have $\alpha=0$ and $F_1(0)=1$, which is exactly 
the unit charge in the current $J^+=\bar\psi\gamma^+\psi$.

\section{Summary}

The advantages of the LFCC method are the absence of Fock-space truncations
that cause uncanceled divergences, elimination of Fock-sector and spectator
dependence of self-energy contributions, and provision for systematic 
improvement, through the addition of terms to the truncated operator $T$.
The terms in $T$ can be organized according to the nature and number
of particles annihilated and created.  Applications to theories beyond
the simple model considered here are in progress, with some preliminary
work on the dressed-electron state in QED already completed~\cite{LFCCqed}.
Additional work in QED will include consideration of the dressed-photon
state, extension of the dressed-electron state to include $e^+$-$e^-$ pairs,
muonium, and positronium.  Application to QCD will begin with consideration
of mesons in holographic QCD~\cite{hQCD}.

\acknowledgments
This work was done in collaboration with 
S.S. Chabysheva and supported in part by
the US Department of Energy and
the Minnesota Supercomputing Institute.

\end{document}